# Expressing Reachability in Linear Recurrences, as Infinite Determinants and Rational Polynomial Equations

Deepak Ponvel Chermakani
deepakc@pmail.ntu.edu.sg   deepakc@e.ntu.edu.sg   deepakc@ed-alumni.net   deepakc@myfastmail.com   deepakc@usa.com

*Abstract: -* We present two tools, which could be useful in determining whether or not a non-Homogenous Linear Recurrence can reach a desired rational. First, we derive the determinant that is equal to the $i^{th}$ term in a non-Homogenous Linear Recurrence. We use this to derive the infinite determinant that is zero, if and only if, the desired rational can be reached by some term in the recurrence. Second, we derive an infinite summation of rational Polynomials, such that this summation can be equal to *1*, if and only if, the desired rational can be reached by some term in the recurrence.

## 1. Introduction

Given a linear recurrence, an interesting question asked by many mathematicians is whether it is possible for zero to be reached by some term in the recurrence (i.e. some term in the recurrence equals zero). This question is still open, and it is not known whether this is decidable [1]. In this paper, we give two tools, which we hope will be useful in answering this Question of Reachability in a non-Homogenous Linear Recurrence. The first tool is an infinite determinant whose value is *0*, if and only if, a desired rational is reachable. The second tool is an infinite summation of rational first degree Polynomials with variable coefficients, such that this summation can become *1*, if and only if, the desired rational is reachable.

Let our non-homogenous linear recurrence be denoted as $E_i$, satisfying the recurrence relation $E_i = f_{i,0} + f_{i,1} E_{i-1} + f_{i,2} E_{i-2} + f_{i,3} E_{i-3} + \ldots + f_{i,L} E_{i-L}$. $i$ denotes Natural Numbers excluding 0, $L$ is a natural number denoting the number of historical values considered for generating the next value of $E_i$, and each of $\{f_{i,0}, f_{i,1}, f_{i,2}, \ldots f_{i,L}\}$ are univariate functions of $i$. Let the first $L$ terms of $E_i$ given to us be $\{E_1 = \alpha_1, E_2 = \alpha_2, E_3 = \alpha_3, \ldots E_L = \alpha_L\}$. Let $r$ be a desired rational, for whom we are trying to determine whether or not atleast one of the terms of recurrence $E_i$ can be equal to $r$ (i.e. also read as whether $E_i$ can reach $r$).

In our paper, if A and B are two Boolean statements, A→B denotes that A is true implies B is true, and A↔B denotes that A is true if and only if B is true.

## 2. Deriving the Infinite Determinant

Let us define a variable $F_0$, and let us define a second linear recurrence $F_i$ where $\{F_1 = \alpha_1 F_0, F_2 = \alpha_2 F_0, F_3 = \alpha_3 F_0, \ldots F_L = \alpha_L F_0\}$, and where $F_i = f_{i,0} F_0 + f_{i,1} F_{i-1} + f_{i,2} F_{i-2} + f_{i,3} F_{i-3} + \ldots + f_{i,L} F_{i-L}$. It now becomes obvious that $(F_0 = 0) \rightarrow (F_i = 0$ for all $i)$. If we were to write out the linear equations for recurrence $F_i$, we have:

$\alpha_1 F_0 - F_1 = 0$
$\alpha_2 F_0 - F_2 = 0$
...
...
$\alpha_L F_0 - F_L = 0$
$f_{L+1,0} F_0 + f_{L+1,L} F_1 + f_{L+1,L-1} F_2 + \ldots + f_{L+1,1} F_L - F_{L+1} = 0$
$f_{L+2,0} F_0 + f_{L+2,L} F_2 + f_{L+2,L-1} F_3 + \ldots + f_{L+2,1} F_{L+1} - F_{L+2} = 0$
...
...
$f_{i,0} F_0 + f_{i,L} F_{i-L} + f_{i,L-1} F_{i-L+1} + \ldots + f_{i,1} F_{i-1} - F_i = 0$
$F_i = 0$

As one of our interests is deciding whether $E_i$ can reach 0, we have placed this as the last equation above. Since all the $F_i$ are expressible as a linear multiple of $F_0$, we can say that $(F_0 \neq 0) \rightarrow ((F_i = 0) \leftrightarrow (E_i = 0))$. Also, setting $F_0 \neq 0$ (for example setting $F_0 = 1$) would imply that a non-trivial solution exists for the vector $< F_0, F_1, F_2, \ldots F_i >$. Hence, from the above set of linear equations, we can derive a determinant $\Omega_i$ that is equal to zero, if and only if, a non-trivial solution exists to the vector $< F_0, F_1, F_2, \ldots F_i >$. In other words $(\Omega_i = 0) \leftrightarrow (E_i = 0)$. The determinant $\Omega_i$ is shown in Figure 1. Due to the presence of 1 in the last column of the last row of the determinant, $\Omega_i$ may be further simplified to the determinant shown in Figure 2.

We now state and prove the following Theorem.

**Theorem-1:** $E_i = \Omega_i$

**Proof:** For this proof, we will refer to the determinant $\Omega_i$ shown in Figure 2. For $i \leq L$, it is easy to see that if $i$ is odd, $\Omega_i = + \alpha_i (-1)^{i-1} = \alpha_i$, and if $i$ is even, $\Omega_i = - \alpha_i (-1)^{i-1} = \alpha_i$. Next, let us focus on calculating $\Omega_i$, for $i > L$. Observe that the last column of this determinant contains two elements -1 and $f_{i,1}$, in the second-last row and last row respectively. If we expand using these two elements, then $\Omega_i = f_{i,1} \Omega_{i-1} - (-1) \gamma_2 = f_{i,1} \Omega_{i-1} + \gamma_2$, where $\gamma_2$ is the determinant obtained by removing the last column and second last row from $\Omega_i$. To expand $\gamma_2$, we again observe the last column of $\gamma_2$ that contains -1 and $f_{i,2}$, so $\gamma_2 = f_{i,2} \Omega_{i-2} + \gamma_3$, and so on until $\gamma_{L-1} = f_{i,L-1} \Omega_{i-L+1} + \gamma_L$, where $\gamma_L$ is the determinant obtained by removing the last $L-1$ columns and the rows numbered $i-L+1$ to $i-1$, from $\Omega_i$. Observe that the last row of $\gamma_L$ has $f_{i,0}$ in the first column and $f_{i,L}$ in the last column. If we were to expand $\gamma_L$ using these two elements, then $\gamma_L = f_{i,L} \Omega_{i-L} + (-1)^{i-L+1} (-1)^{i-L} f_{i,0} = f_{i,L} \Omega_{i-L} + f_{i,0}$. Thus, $\Omega_i = f_{i,1} \Omega_{L-1} + f_{i,2} \Omega_{L-2} + \ldots + f_{i,1} \Omega_{L-1} + f_{i,0}$.

**Hence Proved.**

**Figure 1.** The determinant $\Omega_i$

**Figure 2.** Simplified form of determinant $\Omega_i$

We can now state the following result: ($E_i$ reaches zero) $\leftrightarrow$ (Limit $_{N \to \infty}$ ($\Omega_1 \Omega_2 \ldots \Omega_N$) = 0 provided $\Omega_N$ does not converge to a non-zero Real whose absolute value is lesser than 1). It may be noted that convergence to a non-zero value lesser than $1$ does not occur in Integer Linear Recurrences (i.e. Linear Recurrences where each of $\{f_{i,0}, f_{i,1}, f_{i,2}, \ldots f_{i,L}, \alpha_1, \alpha_2, \alpha_3, \ldots \alpha_L\}$ are integers for all $i$).

More generally, ($E_i$ reaches $r$) $\leftrightarrow$ (Limit $_{N \to \infty}$ (($\Omega_1 - r$)($\Omega_2 - r$) $\ldots$ ($\Omega_N - r$)) = 0 provided ($\Omega_N - r$) does not converge to a non-zero Real whose absolute value is lesser than 1), where $r$ is any desired rational that we would like the linear recurrence to reach. It may be noted that the structure of the determinant whose value = ($\Omega_i - r$) may be obtained by replacing the '0' in the first column of the last row of the determinant mentioned in Figure 1, with '$-r$'. Next, it is known that the product of determinants ($\Omega_1 - r$)($\Omega_2 - r$) $\ldots$ ($\Omega_N - r$) may be represented as a single determinant $\mu$, by placing their structures diagonal to each other. Hence, ($E_i$ reaches $r$) $\leftrightarrow$ (Infinite determinant $\mu = 0$, provided ($\Omega_N - r$) does not converge to a non-zero Real whose absolute value is lesser than 1).

## 3. Deriving the Infinite Summation of degree-one rational Polynomials with variable coefficients

We proceed to our second tool, which might help in determining whether or not a linear recurrence can reach a desired rational. We give the following definitions.

Let the rational expressions $Q_1, Q_2, ... Q_N$ be defined as follows:
$Q_1 = a_1 - 1/x_1$
$Q_2 = a_2 - 1/x_2$
...
$Q_L = a_L - 1/x_L$
$Q_{L+1} = f_{L+1,0} + f_{L+1,L}/x_1 + f_{L+1,L-1}/x_2 + ... + f_{L+1,1}/x_L - 1/x_{L+1}$
$Q_{L+2} = f_{L+2,0} + f_{L+2,L}/x_2 + f_{L+2,L-1}/x_3 + ... + f_{L+2,1}/x_{L+1} - 1/x_{L+2}$
...
$Q_N = f_{N,0} + f_{N,L}/x_{N-L} + f_{N,L-1}/x_{N+1-L} + .. + f_{N,1}/x_{N-1} - 1/x_N$

Let the Polynomials $P_i = (c_{i,1} x_1 + c_{i,2} x_2 + ... + c_{i,N} x_N)$, for all $i$ as integers in $[1,N]$. Here, each $c_{i,j}$ is a real variable for all integers $\{i,j\}$ in $[1,N]$.

We now state and prove the following Theorem.

**Theorem-2**: For all real non-zero values of $x_i$, for all $i$ as integers in $[1,N]$, the following statement is true:
**(Limit $_{N\to\infty}$ ($P_1 Q_1 + P_2 Q_2 + ... + P_N Q_N$) = 1 is allowed) $\leftrightarrow$ (Linear Recurrence $E_i$ reaches 0)**

**Proof**: Consider any system of generic linear equations with an N-dimensional variable vector $\underline{y} = <y_1, y_2 ... y_N>$:
$a_{1,1} y_1 + a_{1,2} y_2 + ... + a_{1,N} y_N = b_1$
$a_{2,1} y_1 + a_{2,2} y_2 + ... + a_{2,N} y_N = b_2$
...
$a_{N,1} y_1 + a_{N,2} y_2 + .. + a_{N,N} y_N = b_N$

Let us denote the constant vector $\underline{a_i} = <a_{1,i}, a_{2,i} ... a_{N,i}>$ consisting of all the constant scalars in column $i$ of Figure 3, and let $b_i$ be some scalar constant, for all integers $i$ in $[1,N]$. Let vector $\underline{b} = <b_1, b_2, ... b_N>$.

Let us further assume that the system of linear equations gives a unique solution. Our assumption is fine, because we are concerned only about modelling the system of linear equations generated by the Linear Recurrence, which obviously always has a unique solution. Hence, the value of the determinant $\Delta$ formed from the matrix of vectors $\underline{a_i}$, shown below in Figure 3, is non-zero. This also means that it is impossible to find a non-trivial constant vector $\underline{v} = <v_1, v_2, ... v_N>$, such that $\underline{v}^T \underline{a_i} = 0$ for all integers $i$ in $[1,N]$. Here we denote $\underline{v}^T$ as denoting the transpose of $\underline{v}$, and $\underline{v}^T \underline{a_i} = a_{1,i} v_1 + a_{2,i} v_2 + ... + a_{N,i} v_N$.

$$\begin{vmatrix} a_{1,1} & a_{1,2} & ... & a_{1,N} \\ a_{2,1} & a_{2,2} & ... & a_{2,N} \\ ... & ... & ... & ... \\ a_{N,1} & a_{N,2} & ... & a_{N,N} \end{vmatrix}$$

**Figure 3.** Determinant formed from Matrix of vectors $\underline{a_i}$

Next, for any integer $t$ in $[1,N]$, the determinant formed by replacing vector $\underline{a_t}$ with $\underline{b}$ in the matrix of vectors $\underline{a_i}$, shown below in Figure 4, is zero, if and only if, $y_t = 0$. This also means that for any integer $t$ in $[1,N]$, there exists a non-trivial constant vector $\underline{w} = <w_1, w_2, ... w_N>$, such that $\underline{w}^T \underline{a_i} = 0$ for all integers $i$ in $[1,N]$ and $i \neq t$, and such that $\underline{w}^T \underline{a_t} = 0$, if and only if, $y_t = 0$.

$$\begin{vmatrix} a_{1,1} & a_{1,2} & ... & a_{1,t-1} & b_1 & a_{1,t+1} & ... & a_{1,N} \\ a_{2,1} & a_{2,2} & ... & a_{2,t-1} & b_2 & a_{2,t+1} & ... & a_{2,N} \\ ... & ... & ... & ... & ... & ... & ... & ... \\ a_{N,1} & a_{N,2} & ... & a_{N,t-1} & b_N & a_{N,t+1} & ... & a_{N,N} \end{vmatrix}$$

**Figure 4.** Determinant formed by replacing vector $\underline{a_t}$ with $\underline{b}$

Now let us replace $y_i$ with $1/z_i$ for all integers $i$ in $[1,N]$, in our generic set of equations, so we have:
$a_{1,1}/z_1 + a_{1,2}/z_2 + ... + a_{1,N}/z_N = b_1$
$a_{2,1}/z_1 + a_{2,2}/z_2 + ... + a_{2,N}/z_N = b_2$
...
$a_{N,1}/z_1 + a_{N,2}/z_2 + .. + a_{N,N}/z_N = b_N$

Let us define $S_i = (a_{i,1}/z_1 + a_{i,2}/z_2 + .. + a_{i,N}/z_N - b_i)$ for all integers $i$ in $[1,N]$.
Let us also define $R_i = (d_{i,1} z_1 + d_{i,2} z_2 + ... + d_{i,N} z_N)$ for all integers $i$ in $[1,N]$. Here, each $d_{i,j}$ is a real variable for all $\{i,j\}$ as integers in $[1,N]$. Let the variable vector $\underline{d_i} = < d_{1,i}, d_{2,i}, ... d_{N,i} >$. In other words, $\underline{d_i}$ represents the coefficients of $z_i$.

From $(R_1 S_1 + R_2 S_2 + ... + R_N S_N)$, collect all coefficients of variables $z_i$ and $z_i/z_j$ for all integers $\{i,j\}$ in $[1,N]$, assuming that $z_i \neq 0$:
Coefficient of 1: $(\sum(\underline{a_i}^T \underline{d_i})$ over $i$ running as integers from $1$ to $N$)
Coefficient of $z_i/z_j$ (for $i \neq j$): $((\underline{a_i}^T \underline{d_j})$ for all integers $\{i,j\}$ in $[1,N]$, and $i \neq j$.
Coefficient of $z_i$: $(- (\underline{d_i}^T \underline{b}))$, for all integers $i$ in $[1,N]$.

We are now in a position to prove the following two Lemmas for any generic system of linear equations with a unique solution.

<u>Lemma-2.1</u>: ($y_t=0$ for some integer $t$ in $[1,N]$) $\rightarrow$ $(((R_1 S_1 + R_2 S_2 + ... + R_N S_N) = 1)$ is allowed for all non-zero real values of $z_i$ for all integers $i$ in $[1,N]$)
<u>Proof</u>: Since $y_t=0$ for some integer $t$ in $[1,N]$, there exists a non-trivial constant vector $\underline{w} = < w_1, w_2, ... w_N>$, such that $\underline{w}^T \underline{a_i} = 0$ for all integers $i$ in $[1,N]$ and $i \neq t$, and such that $\underline{w}^T \underline{b} = 0$. Now let $\underline{d_t} = \underline{w}$. Also, let $\underline{d_i}$ be trivial (i.e. all vector elements are 0) for all integers $i$ in $[1,N]$ and $i \neq t$. Looking back, we see that the coefficients of $z_i/z_j$ and the coefficients of $z_i$ in $(R_1 S_1 + R_2 S_2 + ... + R_N S_N)$ become $0$, for all integers $\{i,j\}$ in $[1,N]$, and $i \neq j$. Since $\underline{d_t}$ is the only non-trivial vector here, the coefficient of 1 in $(R_1 S_1 + R_2 S_2 + ... + R_N S_N)$ becomes equal to $\underline{a_t}^T \underline{d_t}$. Note that $\underline{a_t}^T \underline{d_t} \neq 0$, since we assumed that our linear system of equations have a unique solution. Also, note that the elements of $\underline{d_t}$ are expressible as ratios in terms of each other and hence can be also expressed as a scalar multiple of a single non-zero variable. $\underline{a_t}^T \underline{d_t} = 1$ is then allowed, and so $(R_1 S_1 + R_2 S_2 + ... + R_N S_N) = 1$ is also allowed.
Hence Proved Lemma-2.1

<u>Lemma-2.2</u>: ($y_t \neq 0$ for all integers $t$ in $[1,N]$) $\rightarrow$ $(((R_1 S_1 + R_2 S_2 + ... + R_N S_N) = 1)$ is not allowed for all non-zero real values of $z_i$ for all integers $i$ in $[1,N]$)
<u>Proof</u>: When $y_t \neq 0$ for all integers $t$ in $[1,N]$, it is impossible to find a non-trivial constant vector $\underline{w} = < w_1, w_2, ... w_N>$, such that $\underline{w}^T \underline{a_i} = 0$ for all integers $i$ in $[1,N]$ and $i \neq t$, and such that $\underline{w}^T \underline{b} = 0$. This means that we can only afford to set $\underline{d_i}$ as a trivial vector for all integers $i$ in $[1,N]$, in order to enable the coefficients of $z_i/z_j$ (for $i \neq j$) and $z_i$ vanish in $(R_1 S_1 + R_2 S_2 + ... + R_N S_N)$. But then the coefficient of 1 in $(R_1 S_1 + R_2 S_2 + ... + R_N S_N)$ also vanishes. Thus, $(R_1 S_1 + R_2 S_2 + ... + R_N S_N) = 1$ is not allowed.
Hence Proved Lemma-2.2
**Hence Proved Theorem-2**

Though Theorem-2 is directly meant to help deciding $E_i$'s Reachability to $0$, it can also be applied for Reachability to any rational $r$. This is achieved by a simple replacement of variables in our set of linear equations obtained from the Recurrence. For example, referring to the generic linear equations with a unique solution, '$y_i$' may be written as '$y_i - r + r$', which may be replaced with '$\beta_i + r$', so that $(y_i = r) \leftrightarrow (\beta_i = 0)$. Therefore, the definitions of the rational Polynomials $Q_k$ s may be simply altered as follows: - replace all occurrences of '$1/x_i$' with '$(rx_i+1)/x_i$' in all the $Q_k$ s.

## 4. Conclusion

In this paper, we presented two tools for deciding whether or not a non-homogenous linear recurrence can reach a desired rational. The first tool is an infinite determinant whose value is 0, if and only if, the recurrence reaches the desired rational. As the study of infinite determinants has become popular [2], we hope this tool will be useful here. The second tool is an infinite summation of rational polynomials whose value is allowed to be 1, if and only if, the recurrence reaches the desired rational.

**About the Author**
I, Deepak Ponvel Chermakani, wrote this paper, out of my own interest and initiative, during my spare time. In Sep-2010, I completed a fulltime one year Master Degree course in *Operations Research with Computational Optimization* from University of Edinburgh UK (www.ed.ac.uk). In Jul-2003, I completed a fulltime four year Bachelor Degree course in *Electrical and Electronic Engineering*, from Nanyang Technological University Singapore (www.ntu.edu.sg). I completed my fulltime high schooling from National Public School in Bangalore in India in Jul-1999.